\documentclass{llncs}
\usepackage{graphicx}
\graphicspath{{pics/}{figs/}}
\usepackage{amssymb,latexsym, amsmath, color}
\allowdisplaybreaks

\def\MM#1{\boldsymbol{#1}}
\DeclareMathOperator{\diff}{d}
\DeclareMathOperator{\Diff}{Diff}

\newcommand{\bfi}{\bfseries\itshape}

\newcommand{\trp}[1]{{{\vphantom{#1}}^T#1}}
\newcommand{\rem}[1]{}
\newcommand{\remfigure}[1]{}
\rem{%%%%%%%%%%%%%%%%%%BEGIN REM
\topmargin -10mm
\oddsidemargin 0mm
\evensidemargin 0mm
\textwidth 160mm
\textheight 230mm 
}%%%%%%%%%%%%%%%%%%END REM

\makeatletter
\@addtoreset{equation}{section}

\makeatother

\newcommand{\grid}{grid}

\def\0{{\bf 0}}

\rem{%%%%%%%%%%%%%%%%%%BEGIN REM
\pagestyle{myheadings}
\markright{ Cotter \& Holm 
\hfil 
Singular solutions, momentum maps and computational anatomy
\hfil}
}%%%%%%%%%%%%%%%%%%END REM

\begin{document}

\rem{%%%%%%%%%%%%%%%%%%BEGIN REM
\title{
Singular solutions, momentum maps and computational anatomy}
\author{\vspace{2mm}
C. J. Cotter$^{1}$ and D. D. Holm$^{1,\,2}$\\
$^1$ Department of Mathematics, Imperial College London, London SW7 2AZ, UK\,.\\
$^2$ Computer and Computational Science Division, Los Alamos National Laboratory, 
\\Los Alamos, NM, 87545 USA }
\date{MICCAI  Conference October 1, 2006
}
}%%%%%%%%%%%%%%%%%%END REM

\frontmatter          % for the preliminaries
\pagestyle{empty}  % switches off printing of running heads
\mainmatter              % start of your contributions
\title{Singular solutions, momentum maps and computational anatomy}
\titlerunning{Singular momentum maps and computational anatomy} 
 % abbreviated title (for running head)
%
\author{Colin J. Cotter\inst{1} 
\and Darryl D. Holm\inst{1,2} }
\authorrunning{Cotter and Holm}   % abbreviated author list (for running head)
%
%%%% modified list of authors for the TOC (add the affiliations)
\tocauthor{C. J. Cotter (Imperial College London),
D. D. Holm (Imperial College London and Los Alamos National Laboratory), }
\institute{Department of Mathematics, Imperial College London, 
SW7 2AZ, UK
\email{colin.cotter@imperial.ac.uk}
  \and
  Los Alamos National Laboratory, Los Alamos, NM 87545 USA\\
\email{d.holm@imperial.ac.uk}, \email{dholm@lanl.gov}
%\thanks{We are thankful to XYZ} 
}

\maketitle

\begin{abstract}
This paper describes the variational formulation of template matching problems
of computational anatomy (CA); introduces the EPDiff evolution equation in the
context of an analogy between CA and fluid dynamics; discusses the singular
solutions for the EPDiff equation and explains why these singular solutions
exist (singular momentum map). Then it draws the consequences of EPDiff for
outline matching problem in CA and gives numerical examples.\\

``I shall speak of things . . . so singular in their oddity as in some manner
to instruct, or at least entertain, without wearying." -- Lorenzo da Ponte

\end{abstract}

%\tableofcontents

%%%%%%%%%%%%%%%%
\section{Introduction}

Computational Anatomy (CA) must measure and analyze a range of variations
in shape, or appearance, of highly deformable structures. 
The problem statement for CA was formulated long ago \cite{Th1917}
\begin{quote}
In a very large part of morphology, our essential task lies in the {\bfi
comparison of related forms} rather than in the precise definition of
each\dots. This process of comparison, of recognizing in one form a definite
permutation or deformation of another, \dots  lies
within the immediate province of mathematics and finds its solution in
\dots the Theory of Transformations.
-- D'Arcy Thompson, {\it On Growth and Form} (1917)
\end{quote}
The pioneering work of Bookstein, Grenander and Bajscy
\cite{bajscy-etal-1983,bookstein-1991,grenander-book-1993} first took
up this challenge by introducing a method called template
matching. The past several years have seen an explosion in the use of
template matching methods in computer vision and medical imaging that
is fulfilling D'Arcy Thompson's expectation
\cite{ashburner-etal-2003,GrMi1998,hallinan,jain-etal-1998,miller-etal-2002,miller-younes-2001,montagnat-etal-2001,mumford-shape,mumford-elastica,mumford-pattern,thompson-etal-2000,toga-1999,toga-thompson-2003,Tr1995,Tr1998}. These
methods enable the systematic measurement and comparison of anatomical
shapes and structures in medical imagery.  The mathematical theory of
Grenander's deformable template models, when applied to these
problems, involves smooth invertible maps (diffeomorphisms), as
presented in this context in \cite{miller-etal-2002,miller-younes-2001,Tr1995,Tr1998,DuGrMi1998,mumfordihp}. In particular, the template
matching approach involves Riemannian metrics on the diffeomorphism
group and employs their projections onto specific landmark shapes, or
image spaces.

The problem for CA then becomes to minimize the distance between two images
as specified in a certain representation space, $V$. Metrics are written so
that the optimal path in $V$ satisfies an evolution equation, which was first
discovered in abstract form \cite{arn78} and later called EPDiff when it
arose in the Euler-Poincar\'e theory of optimal motion on smooth invertible
mappings called diffeomorphisms, \cite{HoMa2004}. 

The EPDiff equation coincides with the Euler equation for ideal fluids in the
case that the Riemannian metric for the distance between two images is the
$L^2$ norm. Another type of norm on $V$ (called the $H^1$ norm) arises in the
theory of the fascinating nonlinear coherent solutions of shallow water waves
called {\bfi solitons}. Solitons interact with each other elastically, so they
re-emerge unscathed from fully nonlinear collisions. EPDiff with the $H^1$
norm on $V$ describes the peaked soliton solutions of the Camassa-Holm shallow
water wave equation.  As we shall see, the Camassa-Holm peakons arise from a
general property of Hamiltonian systems called their {\bfi momentum map}. A
discussion of EPDiff and peakons in the particular case of template matching
appears in \cite{HoRaTrYo2004}. 

In this paper, we shall draw parallels between the two
endeavors of fluid dynamics and template matching for computational anatomy, by
showing how the Euler-Poincar\'e theory of ideal fluids can be used to develop
new perspectives in CA. In particular, we discover that CA may be informed by
the concept of weak solutions, solitons and  momentum maps for geodesic flows
\cite{HoRaTrYo2004,CaHo1993,HoSt2004}. 

\subsection{Problem \& Approach for Computational Anatomy}
Computational Anatomy (CA) compares shapes (graphical
structures) by making a {\bfi geodesic deformation} from one shape to the
another. Among these graphical structures, landmarks and image outlines 
in CA are found to be  singular solutions of the geodesic
{\bfi EPDiff} equation. A {\bfi momentum map} for singular solutions of EPDiff
yields their canonical Hamiltonian formulation, which provides
a {\bfi complete parameterization} of the landmarks and image outlines by their
{\bfi canonical positions and momenta}. The momentum map provides an {\bfi
isomorphism} between landmarks (and outlines) for images and {\bfi singular}
(weak) solutions of EPDiff. (These solutions are solitons in 1D.) This
isomorphism provides for CA: 
(1) a complete and non-redundant data representation;
(2) a dynamical paradigm in which image outlines interact
by exchange of momentum;
(3) methods for numerical simulation \& data assimilation.
Euler-Poincar\'e theory also provides a framework for unifying and extending
the various approaches in CA.

Thus, the concept of momentum becomes important for CA, because momentum: 
\\$-\,\,$Completes the representation of images (momentum of cartoons); 
\\$-\,\,$Informs template matching of the possibility of soliton-like collisions and
{\bfi momentum exchange in image outline interactions};
\\$-\,\,$Encodes the subsequent deformation into the
{\bfi initial locus and momentum} of an image outline;
\\$-\,\,$Provides numerical simulation methods using the {\bfi momentum
map for right action} as a data structure; and
\\$-\,\,$Accomplishes matching and data assimilation via the {\bfi adjoint
linear problem} for template matching, using the {\bfi initial
momentum as a control variable}.
\\All of these momentum properties flow from the {\bfi EPDiff}
equation.\smallskip

\noindent
{\bfi Outline of the paper.} Section \ref{CA:sec}
describes the template matching variational problems of computational
anatomy, explains the analogy with fluid dynamics and introduces the
fundamental EPDiff evolution equation.  The singular solutions for the EPDiff
equation (\ref{EP-eqn}) with diffeomorphism group $G$ are discussed in section
\ref{Outlines:sec}. They are, in particular, related to the outline matching
problem in computer vision,  examples of which are given in section
\ref{numerex:sec}.

\section{Mathematical formulation of template matching for CA}
\label{CA:sec}

\subsection{Cost}

Most problems in CA can be formulated as: {\bfi Find the deformation path
(flow) with minimal cost, under the constraint
  that it carries the template to the target}. Such problems have a remarkable
analogy with fluid dynamics. The {\bfi cost} assigned in template matching for 
comparing images $\mathcal{I}_0$ \& $\mathcal{I}_1$ is specified as a
functional 
\[
\text{Cost}(t\mapsto\varphi_t) = \int_0^1 \ell({\mathbf{u}}_t) \,dt
\]
defined on curves $\varphi_t$ in a Lie group with tangents
$%\begin{equation}\label{eq:1}
\frac{d\varphi_t}{dt} = {\mathbf{u}}_t \circ \varphi_t
$
and 
$\mathcal{I}_t = \varphi_t\cdot \mathcal{I}_0
.$ %\end{equation}
In what follows, the function ${\mathbf{u}}_t\mapsto
\ell({\mathbf{u}}_t)=\|{\mathbf{u}}_t\|^2$ will be taken as a
squared functional norm on the space of velocity vectors along the flow. 
The Lie group property specifies the representation space for template matching
as a manifold of smooth mappings, which may be differentiated, composed and
inverted. The vector space of {\bfi right invariant}
instantaneous velocities,
${\mathbf{u}}_t=(d\varphi_t/dt)\circ\varphi_t^{-1}$ forms the tangent space
at the identity of the considered Lie group, and may be identified as the
group's  {\bfi Lie algebra}, denoted {$\mathfrak{g}$}.

\subsection{Mathematical analogy between template matching and
fluid dynamics}

(I) The frameworks of CA and fluid dynamics both involve a {\bfi
right-invariant} stationary principle with action, or cost function. The main
differences are that template matching is formulated as an optimal control
problem whose cost function is designed for the application, while fluid
dynamics is formulated as an initial value problem whose cost function is the
fluid's kinetic energy.

\noindent
(II) The geodesic evolution for both template matching and fluid
dynamics is governed by the {\bfi EPDiff equation}
\cite{HoMaRa1998,mumfordihp},
\begin{equation}\label{EP-eqn}
\Big(\frac{\partial}{\partial t}
+ {\mathbf{u}}\cdot\nabla\Big)
{\mathbf{m}}
+\,
(\nabla {\mathbf{u}})\trp\cdot\,{\mathbf{m}}
+
{\mathbf{m}}({\rm div\,}{\mathbf{u}})
=
0
\,.
\end{equation}
Here ${\mathbf{u}}=G*{\mathbf{m}}$, where $G*$ denotes convolution
with the Green's kernel $G$ for the operator $Q_{op}$, where
\[
{\mathbf{m}} = 
\frac{\delta \ell}{\delta \mathbf{u}}
=:
Q_{op}\mathbf{u}
\]
The operator $Q_{op}$ is symmetric and positive definite for the cost defined
by
$$
\text{Cost}(t\mapsto\varphi_t) = \int_0^1 \ell({\mathbf{u}}_t) \,dt
 = \frac{1}{2}\int_0^1
\|{\mathbf{u}}_t\|^2\,dt
=
\frac{1}{2}\int_0^1 \langle \,\mathbf{u}_t\,,\,Q_{op}\mathbf{u}_t\,\rangle\,dt
$$
with $L^2$ pairing $\langle \,\cdot\,,\,\cdot\,\rangle$ whenever
$\|{\mathbf{u}}_t\|^2$ is a norm. 

\noindent
(III) The flows in CA and fluid dynamics both evolve under a left group
action on a linear representation space,
$
\mathcal{I}_t = \varphi_t\cdot \mathcal{I}_0
.$
They differ in the roles of their advected quantities, $a_t=a_0\circ
\varphi_t^{-1}$. The main difference is that image properties are
passive and affect the template matching as a constraint in the cost
function, while advected quantities may affect fluid
flows directly, for example through the pressure, so as to produce waves.

%%%%%%%%%%%%%%%%%%%%%%%%%%%%%%%%%%%%%%%%%%%%%%%%%%%%%%%%%%%%%%%%%%%%%%%%
\subsection{How EPDiff emerges in CA} 

Choose the cost function for
continuously morphing 
$\mathcal{I}_0$ into
$\mathcal{I}_1$ as
\[
\text{Cost}(t\mapsto\varphi_t) 
= \int_0^1 \ell({\mathbf{u}}_t) \,dt
= \int_0^1
\|{\mathbf{u}}_t\|^2\,dt
\,,
\]
where $u_t$ is the velocity of the fluid
deformation at time $t$ and \[\|{\mathbf{u}}_t\|^2 =
\langle \,\mathbf{u}_t\,,\,Q_{op}\mathbf{u}_t\,\rangle\,,\] 
and $Q_{op}$ is our positive symmetric linear
operator.  Then, the momentum governing the process,
$\mathbf{m}_t = Q_{op}\mathbf{u}_t$
with Green's function $G:\,\mathbf{u}_t=G*\mathbf{m}_t$
satisfies the EPDiff equation, (\ref{EP-eqn}).
\rem{
\[
\Big(\frac{\partial}{\partial t}
+ {\mathbf{u}}\cdot\nabla\Big)
{\mathbf{m}}
+\,
(\nabla {\mathbf{u}})\trp\cdot{\mathbf{m}}
+
{\mathbf{m}}({\rm div\,}{\mathbf{u}})
=
0
\]
}
This equation arises in both template matching and fluid dynamics, and it
informs both fields of endeavor. 

%%%%%%%%%%%%%%%%%%%%%%%%%%%%%%%%%%%%%%%%%%%%%%%%%%%%%%%%%%%%%%%%%%%%%%%%
\subsection{Deriving EPDiff from Euler-Poincar\'{e} Reduction of Hamilton's
principle}.

{\bfi Euler-Poincar\'{e} Reduction\/} starts with a {right} (or left)
$G-$invariant Lagrangian $L:TG\rightarrow\mathbb{R}$ on the tangent bundle of
a Lie group $G $. {\bfi Right invariance} of
the Lagrangian may be written as 
\[L(g(t) ,\dot g(t))=L(g(t)h ,\dot g(t)h)
\,,
\hbox{ for all }h\in G
\]
A $G-$invariant Lagrangian defined on $TG$
possesses a symmetry-reduced Hamilton's principle defined on the Lie algebra
$TG/G\simeq\mathfrak{g}.$
%\[TG\mapsto TG/G\simeq\mathfrak{g}\] 
Stationarity of the symmetry-reduced Hamilton's
principle yields the {\bfi Euler-Poincar\'e equations} on the dual Lie algebra
$\mathfrak{g}^*$.  For $G=\Diff$, this equation is {\bfi EPDiff}  (\ref{EP-eqn}).

%%%%%%%%%%%%%%%%%%%%%%%%%%%%%%%%%%%%%%%%%%%%%%%%%%%%%%%%%%%%%%%%%%%%%%%%
\section{Outline matching \& momentum measures}\label{Outlines:sec}

{\bfi Problem statement for outline matching:}\\
Given two collections of curves $c_1,
\ldots, c_N$ and $C_1, \ldots, C_N$ in $\Omega$, find a time-dependent
diffeomorphic process
$(t\mapsto \varphi_t)$ of minimal action (or cost) such that $\varphi_0 =
\text{id}$ and
$\varphi_1(c_i) = C_i$ for $i= 1, \ldots, N$. The matching problem for the
image outlines  seeks {\bfi singular momentum solutions} which naturally
emerge in the computation of geodesics. 
%%%%%%%%%%%%%%%%%%%%%%%%%%%%%%%%%%%%%%%%%%%%%%%%%%%%%%%%%%%%%%%%%%%%%%%%

%%%%%%%%%%%%%%%%%%%%%%%%%%%%%%%%%%%%%%%%%%%%%%%%%%%%%%%%%%%%%%%%%%%%%%%%
\subsection{Image outlines as Singular Momentum Solutions of
EPDiff}

For example, in the 2D plane, EPDiff has weak
{\bfi singular momentum solutions} that are expressed as
\cite{CaHo1993,HoMa2004,HoSt2004}
\begin{equation}\label{EP-sing-mom}
{\mathbf{m}}({\mathbf{x}},t)
=
\sum_{a=1}^N\int_{s}
{\mathbf{P}}_a(t,s)\delta\big({\mathbf{x}}-{\mathbf{Q}}_a(t,s)\big)\,ds
\,,
\end{equation}
where $s$ is a {\bfi Lagrangian coordinate} defined along a set of
$N$ curves in the plane {\it moving with the flow} by the equations
${\mathbf{x}}={\mathbf{Q}}_a(t,s)$ and supported on the delta functions in
the EPDiff solution (\ref{EP-sing-mom}). Thus, the singular momentum solutions
of EPDiff represent evolving ``wavefronts'' supported on delta functions
defined along curves ${\mathbf{Q}}_a(t,s)$ with arclength coordinate $s$ and
carrying momentum ${\mathbf{P}}_a(t,s)$ at each point along the curve as
specified by (\ref{EP-sing-mom}). These solutions exist in any dimension
and they provide a means of performing CA matching for points (landmarks),
curves and surfaces, in any combination.

%%%%%%%%%%%%%%%%%%%%%%%%%%%%%%%%%%%%%%%%%%%%%%%%%%%%%%%%%%%%%%%%%%%%%%%%
\subsection{Here is the Geometry -- Leading to the Numerics}
The basic observation that ties everything together in $n-$dimensions is the
following:\\
{\bfi
Theorem (Holm and Marsden, \cite{HoMa2004}): EPDiff singular momentum solutions
$
T^*{\rm Emb}(S,\mathbb{R}^n)\to\mathfrak{g}^*: \
(\mathbf{P},\mathbf{Q})\to\mathbf{m}
$
define a momentum map.}

It is beyond our scope here to explain either the proof
of this theorem or the mathematics underlying momentum maps for
diffeomorphisms. However, we summarize the main results for template matching, as
follows:
\\$-\,\,$
The embedded manifold $S$ is the support set of the $P$'s and $Q$'s.
\\$-\,\,$
The momentum map is for left action of the diffeomorphisms on $S$.
\\$-\,\,$
The whole system is right invariant.
\\$-\,\,$
Consequently, its momentum map for right action is conserved.
\\$-\,\,$
These constructions persist for a certain class of numerical schemes.
\\$-\,\,$
They apply in template matching for every choice of norm.

%%%%%%%%%%%%%%%%%%%%%%%%%
\subsection{A familiar example of a momentum map}
%%%%%%%%%%%%%%%%%%%%%%%%%%%%%%%%%%%%%%%%%%

A {\bfi momentum map} $\mathbf{J}:\,T^*Q\mapsto\mathfrak{g}^*$ is a 
Hamiltonian for the canonical action of a Lie group $\sf{G}$ on phase space
$T^*Q$. It is expressed in
terms of the pairing
$\langle\,\cdot\,,\,\cdot\,\rangle: 
\mathfrak{g}^*\times\mathfrak{g}\mapsto\mathbb{R}$
as 
\[
\langle\,\mathbf{J}\,,\,\xi\,\rangle
=
\langle\,p\,,\,\pounds_\xi q\,\rangle
=:
\langle\,q\diamond p\,,\,\xi\,\rangle
\,,
\]
where $(q,p)\in T_q^*Q$ and the Lie derivative $\pounds_\xi q$ is the
infinitesimal generator of the action of the Lie algebra element
$\xi\in\mathfrak{g}$ on $q$ in the manifold $Q$. 

The standard example is 
$\pounds_\xi q=\xi\times q$ for
$\mathbb{R}^3\times\mathbb{R}^3\mapsto\mathbb{R}^3$, with pairing
$\langle\,\cdot\,,\,\cdot\,\rangle$ given by scalar product of vectors. The
momentum map is then
\[
\mathbf{J}\cdot \xi
=
p \cdot \xi\times q 
= 
q \times p \cdot \xi
\Rightarrow \mathbf{J} = q \times p\]  
This is {\bfi angular momentum}, the Hamiltonian for phase-space rotations. 
The outlines of images may be parameterized as curves whose dynamics must be
invariant under reparameterizing the arclengths that label those curves. This
symmetry leads to a conserved momentum map, called the {\bfi circulation}
along the curves. The analog of this conservation law for fluids is the
classical Kelvin circulation theorem. 

%%%%%%%%%%%%%%%%%%%%%%%%%%%%%%%%%%%%%%%%%%%%%%%%%%%%%%%%%%%%%%%%%%%%%%%%
\subsection{EPDiff dynamics informs optimal control for CA}

CA  must compare two geometric objects, and thus it is concerned with 
 an {\bfi optimal control problem}. However, the
 {\bfi initial value problem} for EPDiff also has
{\bfi important consequences for CA applications}.

\begin{itemize}
\item
When
matching two geometric structures, the {\bfi momentum at time {\it t=0}
contains all required information for reconstructing the target from the
template}. This is done via {\bfi Hamiltonian geodesic flow}.
\item
Being canonically conjugate, the momentum has exactly the same dimension as the
matched structures, so there is {\bfi no redundancy}. 
\item
Right invariance mods out the relabeling motions from the optimal solution.
This symmetry also yields a {\bfi conserved momentum map}.

\item
Besides being one-to-one, the momentum
representation is defined on a {\bfi linear space}, being dual to the
velocity vectors. \smallskip

This means one may, for example,:
\begin{itemize}
\item
study linear instability of CA processes, 
\item
take averages and 
\item
apply statistics to the space of image contours. 
\end{itemize}
The {\bfi advantage} is the ease of building, sampling and
estimating statistical models on a {\bfi linear space}.

\rem{
\begin{enumerate}
\item
One may apply {\bfi linear perturbations} of either velocity fields, or
momenta to the template. 
\item
The {\bfi average} of a collection of
momenta, of their principal components, or time derivatives of momenta at
a fixed template are all well-defined quantities.
\item
Any {\bfi statistical model based on momentum} provides, through EPDiff, a
statistical model on the deformations,
\item
The {\bfi advantage} is the ease of building, sampling and
estimating statistical models on a {\bfi linear space}. 
\end{enumerate}
}

\end{itemize}

%%%%%%%%%%%%%%%%%%%%%%%%%%%%%%%%%%%%%%%%%%%%%%%%%%%%%%%%%%%%%%%%%%%%%%%%
\subsection{Summary}\label{conc-sec}

We have identified {\bfi momentum as a key concept} in the
representation of image data for CA and discussed analogies with fluid
dynamics. The {\bfi fundamental idea} transferring from fluid dynamics to CA
is the idea of {\bfi momentum maps} corresponding to group
actions.

%%%%%%%%%%%%%%%%%%%%%%%%%%%%%%%%%%%%%%%%%%%%%%%%%%%%%%%%%%%%%%%%%%%%%%%%
\section{Numerical examples of outline matching}\label{numerex:sec}
%%%%%%%%%%%%%%%%%%%%%%%%%%%%%%%%%%%%%%%%%%%%%%%%%%%%%%%%%%%%%%%%%%%%%%%%

In this section we describe our new technique applying
particle-mesh methods to the problem of matching outlines. First we
describe the approach to calculating geodesics in the space of
outlines.

Let $\MM{Q}_0$ and $\MM{Q}_1$ be two embeddings of $S^1$ in
$\mathbb{R}^2$ which represent two shapes, each a closed planar curve.
We seek a 1-parameter family of embeddings
$\MM{Q}(t):S^1\times[0,1]\to\mathbb{R}^2$ so that $\MM{Q}(0)=\MM{Q}_0$
and $\MM{Q}(1)$ matches $\MM{Q}_1$ (up to relabeling). $\MM{Q}(t)$ is
found by minimizing the constrained norm of its velocity.  To find the
equation for $\MM{Q}$ we require extremal values of the action
\[
A = \!\!\int_0^1 \frac{1}{2} L(\MM{u})
\diff{t}
+
 \!\!\int_0^1\!\!\!\int_{S^1}\MM{P}(s,t)
\cdot(\MM{\dot{Q}}(s,t)-\MM{u}(\MM{Q}(s,t)))\diff{t},
\qquad L = \|\MM{u}(t)\|_\mathfrak{g}^2,
\]
\emph{i.e.} we seek time-series of vector fields $\MM{u}(t)$ which are
minimized in some norm subject to the constraint that $\MM{Q}$ is
advected by the flow using the Lagrange multipliers $\MM{P}$ (which we
call momentum). The minimizing solutions are
\begin{eqnarray}
\frac{\delta L}{\delta \MM{u}}
& = & \int_{S^1}\MM{P}(s,t)\delta(\MM{x}-\MM{Q}(s,t))\diff{s},
\label{mom map curve}
\\
\MM{\dot{P}}(s,t) & = & -\,\MM{P}(s,t)\cdot\nabla\MM{u}(\MM{Q}(s,t),t), 
\\
\MM{\dot{Q}}(s,t) & = & \MM{u}(\MM{Q}(s,t),t), \label{mom map curve 2}
\end{eqnarray}
subject to $\MM{Q}(s,0)=\MM{Q}_0(s)$.

We note that equation (\ref{mom map curve}) is the momentum map
corresponding to the cotangent-lift of the action of vector fields
$\MM{u}$ on embedded curves given by
\[
\MM{Q} \mapsto \MM{u}(\MM{Q}).
\]
For a suitable test function $\MM{w}$, we obtain
\[
\frac{d}{dt}\langle\MM{w},\MM{m}\rangle
- \langle \nabla\MM{w},\MM{u}\MM{m}\rangle
+ \langle \MM{w},(\nabla\MM{u})^T\cdot\MM{m}\rangle=0,
\qquad \MM{m} = \frac{\delta L}{\delta \MM{u}},
\]
which is the weak form of the EPDiff equation.

Now one must seek initial momentum $\MM{P}(s,0)$ which takes shape
$\MM{Q}_0(s)$ to shape $\MM{Q}_1(s)$. To do this, we choose some 
functional $J$ of the advected shape $\MM{Q}(1,s)$ which is minimized
when $\MM{Q}(1,s)$ matches $\MM{Q}_1(s)$. Following \cite{GlTrYo04},
we describe the curves by singular densities:
\begin{eqnarray}
\label{singular densities 1}
\mu &=&
\int_{S_1}\hat{\mu}(s)\delta(\MM{x}-\MM{Q}(1,s)\diff{s}\diff{V}(\MM{x}), \\
\label{singular densities 2}
\qquad
\eta &=& \int_{S_1}\hat{\eta}(s)\delta(\MM{x}-\MM{Q}_1(s)\diff{s}\diff{V}(\MM{x}),
\end{eqnarray}
and write $J = \|\mu-\eta\|^2_G$ where $\|\cdot\|^2_G$ is a norm for a
densities in a reproducing kernel Hilbert space with kernel $G$. This
approach means that we do not need to force particular points to be
matched to each other on the shapes. This last problem can be solved
by using a gradient algorithm, where the gradient of the residual
error with respect to $\MM{P}(s,0)$ is calculated using the adjoint
equation \cite{Gu03}.

%%%%%%%%%%%%%%%%%%%%%%%%%%%%%%%%%%%%%%%%%%%%%%%%%%%%%%%%%%%%%%%%%%%%%%%%
\subsection{Numerical discretization}
We use the Variational Particle-Mesh (VPM) method
\cite{Co2005,CoHo2006} to discretize the equations (\ref{mom map
curve}-\ref{mom map curve 2}), as follows: discretize the velocity on
an Eulerian grid with $n_g$ points and approximate $\|\MM{u}\|$ there;
replace $S^1$ by representing the shape by a finite set of $n_p$
Lagrangian particles $\{\MM{Q}_\beta\}_{\beta=1}^{n_p}$, and
interpolate from the grid to the particles using basis functions
\[
\MM{u}(\MM{Q}_\beta) = \sum_{k=1}^{n_g}\MM{u}_k\psi_k(\MM{Q}_\beta)
\,,\quad\hbox{with}\quad
\sum_{k=1}^{n_g}\psi_k(\MM{x}) = 1
\,,\ \forall\ \MM{x}.
\]
The action for the continuous time motion on the grid then becomes
\[
A = \int_0^1 \frac{1}{2}\|\MM{u}(t)\|_{\grid}^2+\sum_\beta\MM{P}_\beta
\cdot\left(\MM{\dot{Q}}_\beta-\sum_k\MM{u}_k\psi_k(\MM{Q}_\beta)\right)\diff{t},
\]
and one can obtain a fully discrete method by discretizing the action
in time. For example, we can obtain a first-order method by extremizing
\[
A = \Delta
t\sum_{n=1}^N\left(\frac{1}{2}\|\MM{u}^n\|_{\grid}^2+\sum_\beta\MM{P}^n_\beta
\cdot\left(\frac{\MM{Q}^n_\beta-\MM{Q}^{n-1}_\beta}{\Delta t}
-\sum_k\MM{u}_k^n\psi_k(\MM{Q}^{n-1}_\beta)\right)\right).
\]
The resulting time-stepping method
\rem{\begin{eqnarray*}
\MM{m}_k^n = \nabla_{\MM{\MM{u}_k^n}} \frac{1}{2}\|\MM{u}^n\|_{\grid}^2
& = & \sum_\beta\MM{P}^n_\beta
\psi_k(\MM{Q}^{n-1}_\beta), \\
\MM{P}^{n} & = & \MM{P}^{n-1} - \Delta t
\MM{P}^n\cdot\sum_k\MM{u}^n_k\nabla\psi_k(\MM{Q}_\beta^{n-1}), \\
\MM{Q}^n & = & \MM{Q}^{n-1} + \Delta t\sum_k\MM{u}_k\psi_k(\MM{Q}_\beta^{n-1}),
\end{eqnarray*}}
is the (first-order) symplectic Euler-A method for the time-continuous
Hamiltonian system for the Lagrangian particles.  In general, the
method will always be symplectic since it arises from a discrete
variational principle (see \cite{LeRe2005} for a broad introduction to
symplectic numerical methods and their conservation properties). The
conservation properties of VPM are discussed in \cite{CoHo2006}.

\begin{figure}[htp]
\begin{center}
\scalebox{0.25}{\includegraphics{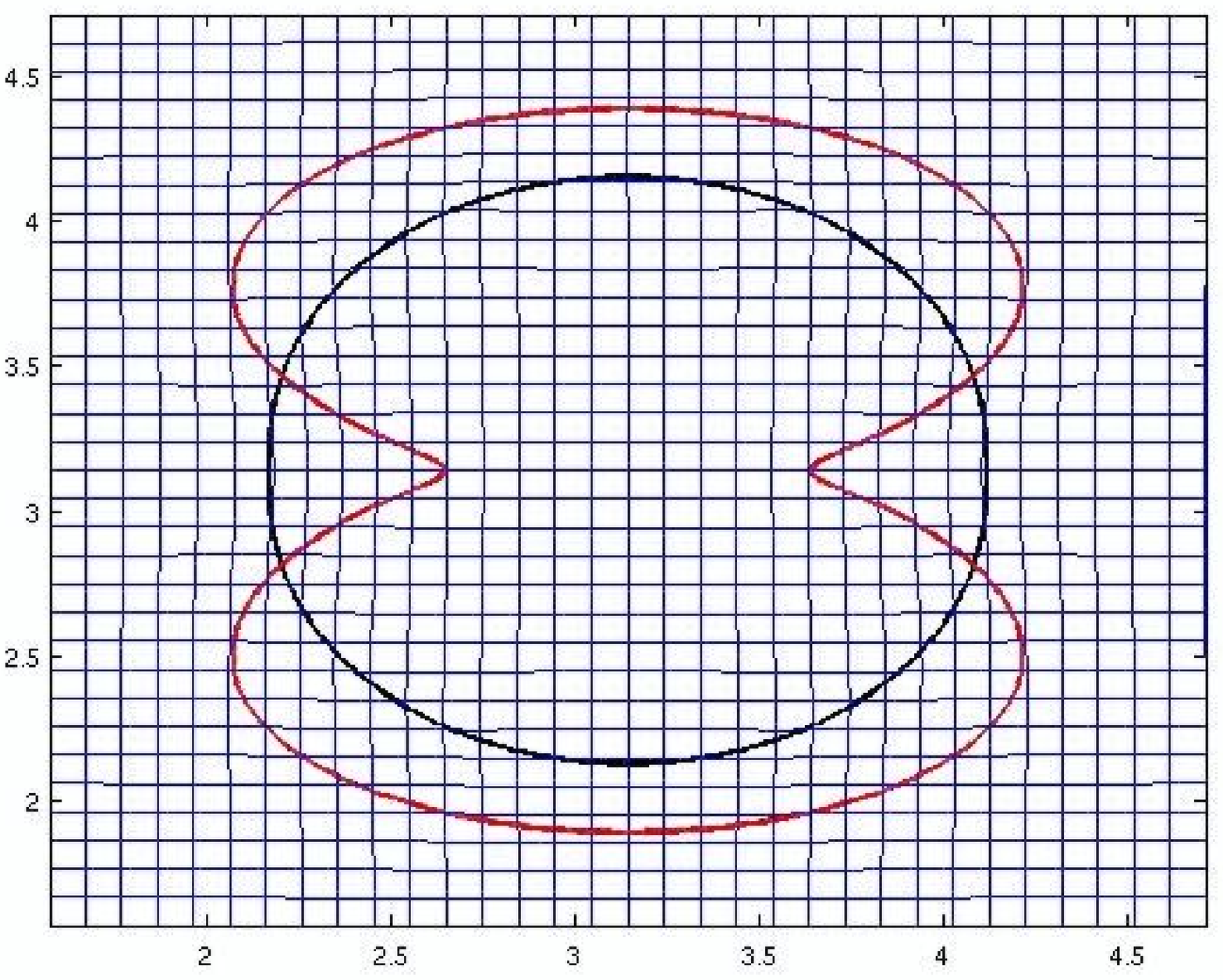}}\scalebox{0.25}{\includegraphics{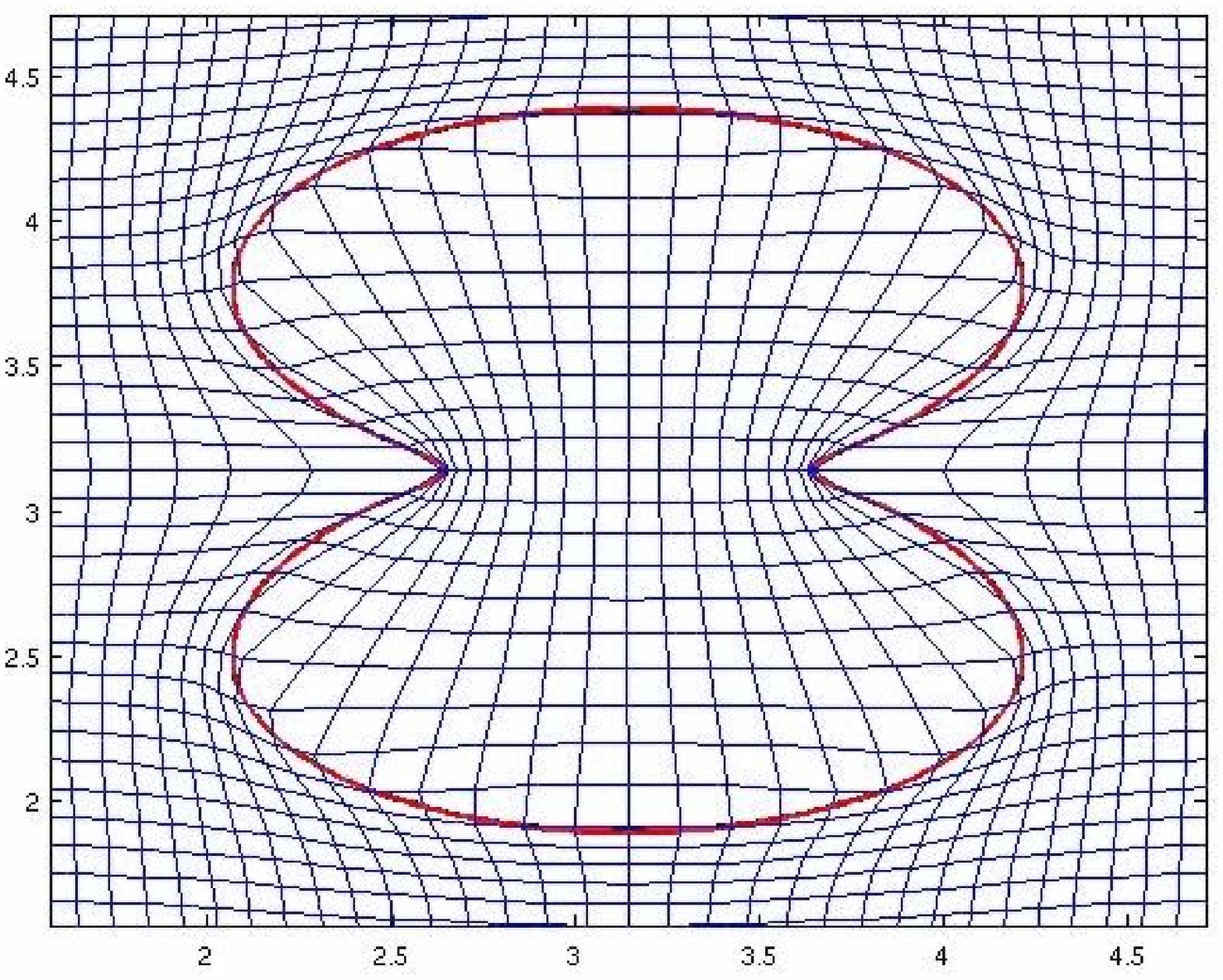}}
\end{center}
\caption{\label{cusp} Results from a VPM calculation to calculate the
minimal path between a two simple shapes. On the left, the initial and
final shapes are shown, and on the right, the deformation of the
initial shape into the final shape is depicted together with a grid
which shows how the flow map deforms the space around the shape. We
used the $H^1$ norm for velocity on a $2\pi\times2\pi$ periodic domain
on a $128\times128$ grid, discretized using FFT, and the corresponding
kernel was used to calculate $J$. Cubic B-splines were used as basis
functions.}
\end{figure}

We approximate the densities $\mu$ and $\eta$ on the grid using the standard
particle-mesh approach (see \cite{FrGoRe2002}):
\[
\mu_k = \sum_\beta\hat{\mu}_\beta\psi_k(\MM{Q}^N_\beta), \qquad
\eta_k = \sum_\beta\hat{\eta}_\beta\psi_k(\MM{Q}_{1,\beta}),
\]
where $\MM{Q}_{1,\beta}$ are the positions of particles on the target
shape. This amounts to ``pixellating'' the singular densities
(\ref{singular densities 1}-\ref{singular densities 2}) on the
grid. For a given kernel $G$, we approximate $J$ with
\[
J = \sum_{kl}G(\MM{x}_k-\MM{x}_l)(\mu_k-\eta_k)(\mu_l-\eta_l).
\]
The discrete adjoint is then applied in computing the inversion for
the initial conditions for $\MM{P}_\beta$ which generate the flow. A 
numerical example calculated using this method is given in figure
\ref{cusp}.

%%%%%%%%%%%%%%%%%%%%%%%%%%%%%%%%%%%%%%%%%%%%%%%%%%%%%%%%%%%%%%%%%%%%%%%%
\subsection*{Acknowledgments}  
This work was partially supported by US DOE, under contract
W-7405-ENG-36 for Los Alamos National Laboratory, and Office of
Science ASCAR/AMS/MICS. 

%%%%%%%%%%%%%%%%%%%%%

\bibliographystyle{splncs}
\bibliography{miccai}

%\end{multicols}%{2}
\end{document}